\documentclass[11pt,a4paper]{article}
\pdfoutput=1

\usepackage{amsmath}
\usepackage{amsfonts}
\usepackage{amsthm}
\usepackage{amssymb}
\usepackage{amscd}
\usepackage[british]{babel}
\usepackage{graphicx}
\usepackage{psfrag}
\usepackage{epsfig}
\usepackage{rotating}
\usepackage{times}
\usepackage{color}
\usepackage{ulem}
\usepackage{longtable}

\theoremstyle{plain}

\newcommand{\boxend}{\flushright{$\Box$}}

\begin{document}

\title{The twilight of the single field slow rolling inflaton}

\author{Jaume Amor\'os$^{a,}$\footnote{E-mail: jaume.amoros@upc.edu} and Jaume de Haro$^{a,}$\footnote{E-mail: jaime.haro@upc.edu}}

\maketitle

\begin{center}
{\small
$^a$Departament de Matem\`atica Aplicada I, Universitat
Polit\`ecnica de Catalunya \\ Diagonal 647, 08028 Barcelona, Spain \\
}
\end{center}

\thispagestyle{empty}

\begin{abstract}
We present a study of 49 single field, slow roll inflationary potentials
in which we assess the likelyhood of these models fitting the spectral
parameters
of the CMB radiation, namely the spectral index $n_s$, its
running $\alpha_s$, the running of the running $\beta_s$ and the ratio of tensor to scalar perturbations $r$,  according
to the currently most accurate determination of these spectral parameters
given by the PLANCK collaboration. A double, partly redundant approach is followed:
for most models we derive analytically bounds for the
values of the spectral parameters that they can support, and
for all of the models we check numerically with a MATLAB program
the spectral parameters that each model can yield for a very broad,
comprehensive list of possible parameter and field values.

The comparison of spectral parameter values supported by the models
with their successive determinations by the PLANCK collaboration
leads to contradictory conclusions: PLANCK2013+WP+BAO:$\Lambda$CDM+$r$+$\alpha_s$
disfavours all of the models with confidence at least 93\%, conversely the data provided by PLANCK2015+TT+lowP
without running of the running allows back most of the models, but taking into account the running of the running again disfavours
39 of the 49 models with confidence at least 92.8\%, and 5 more
because of the amount of expansion e-folds supported.

We identify a bias
in the method of determination of the spectral parameters currently
used to reconstruct the power spectrum of scalar perturbations
that can explain these contradictory conclusions. The solution to this problem is likely to determine the fate of the inflaton.
\end{abstract}


\maketitle

\section{Introduction}
The PLANCK2013 data \cite{Ade} showed that the spectral index of scalar perturbations has an expected negative running, whose modulus was in general one order of magnitude greater than
the theoretical values
provided by slow roll inflation,
which could be used to test some theoretical slow roll inflationary models.
The aim of the present work is to present an analytic and numerical study, of the 49 theoretical models of single field, slow roll inflation that are described in \cite{MARTIN}
(see its Table 1),
that shows that the constrains of the spectral index, its running and the ratio of tensor to scalar perturbations provided by PLANCK2013 combined with several data, disfavours
all of these models, severely in most cases.
The most familiar ones such as Hill-top, Natural, Plateau and Monomial potentials could be disregarded according to these PLANCK2013 data due to the small value, in modulus, of the theoretical value of the running
provided by slow roll inflation. In fact,
from  PLANCK2013+WP+BAO:$\Lambda$CDM+$r$+$\alpha_s$  data
we show that the deviation from the theoretical value of the running, namely ${\frak D}$, obtained from the majority of that models
to its expected observational value is larger than $1.6\sigma$, and in  modulus
 all the theoretical values
are smaller than the modulus of the expected observational one, which means that  these models lies outside of the 94.5\% C.L. Further numerical study has shown that for all
of the models either $\alpha_s$ or $n_s$ or $r$ lie outside of the 93\% C.L.

Our study, initially performed from PLANCK2013 data,  could suggest that single scalar inflationary theories must be replaced by multiple fields theories,
 by other ones with a breakdown of the slow roll phase \cite{running}
or by reconstruction techniques \cite{Odintsov}.
Another completely different proposal is to abandon the inflationary paradigm in favour of the Matter Bounce Scenario \cite{cai,Brandenberger,Haro}
where the Big Bang singularity is replaced by a non-singular Big Bounce,
which at this moment,  constitutes a promising alternative to the slow roll inflationary paradigm, without the characteristic
inflationary flaws,
such as the initial singularity which  or the fine-tuning of the degree of flatness required for the potential in order to achieve successful inflation \cite{ffg91}.

Fortunately for single field slow roll inflation, the new PLANCK2015 observational data \cite{Planck:2015xua}, reduce the modulus of the running one order, which allows the viability
of the majority of single scalar field slow roll inflationary models. For example, using
Planck2015+TT+lowP:$\Lambda$CDM+$r$+$\alpha_s$, where $n_s=0.9667\pm 0.0066$
and $\alpha_s=-0.0126^{+0.0098}_{-0.0087}$
at $1\sigma$ C.L. (see table $4$ of \cite{Planck:2015xua}), and the conservative constrain $r\leq 0.25$
(see figure $6$ of \cite{Planck:2015xua}),
for the majority of tested potentials  one has ${\frak D}\cong 1.1\sigma$, which means that
these models are only disfavored at $86$\% C.L. or less. Moreover, if one introduces lensing, then ${\frak D}$ will be reduced to be lower than $1\sigma$, and thus allowing
single scalar field slow roll
inflation.

However, dealing with the running of the running, namely $\beta_s$,  when one considers the  PLANCK2015 TT+lowP (resp. PLANCK2015 TT,TE,EE+lowP) model
$\beta_s=0.029^{+0.015}_{-0.016}$
(resp. $\beta_s=0.025\pm 0.0013$) (see (19) of \cite{Planck:2015xua}), the same
methodology that we have applied in the previous cases show that the measured value of
$\beta_s$ is incompatible with the theoretical value
provided by single field slow roll,
because its measured value is too large for a magnitude that depends on third order on the slow roll parameters, disfavouring almost all slow roll inflationary models studied,
39 out of 49 at 92.8 \% C.L. or more.

\medskip

The main lesson of this work,  is the importance of the observational measures provided by PLANCK or other teams, in order to check the viability of single field slow roll inflation. Only after
a precise determination of their
expected observational value and the corresponding deviation
we will able to determine which inflationary models could depict correctly our Universe.

The problem of the reliability of the observational measures
of spectral parameters of the CMB starts with the above paradox of the the oscillating conclusions, from ruling
out every model to allowing most of them and back, that the successive PLANCK2013 and PLANCK2015 determinations of spectral parameters support. We find that these oscillating
conclusions can be explained by a bias in the method of fitting spectral values of the power spectrum to the
observations. We believe that new data, and the addressing of this bias, will reduce significantly the observable value
of the running of the running and may possibly allow back
the viability of single slow roll inflation, in the same way as has happened with the early
values of the running obtained by PLANCK2013 and its drastic decrease according to
PLANCK2015.

\medskip

The units used in the paper are: $\hbar=c=8\pi G=1.$

\medskip

\section{ Slow-roll parameters} \label{s:formules}
In  slow roll inflation (see \cite{Basset} for a review of inflation) the commonly used  first order parameters are:
\begin{eqnarray}\label{x14}
\epsilon= -\frac{\dot{H}}{H^2}\cong \frac{1}{2}\left(\frac{V_{\varphi}}{V} \right)^2 \quad\mbox{and}\quad \eta=2\epsilon-\frac{\dot{\epsilon}}{2H\epsilon}
\cong\frac{V_{\varphi\varphi}}{V}.
\end{eqnarray}

At the first slow roll order,
the spectral index  of scalar perturbations and its running are given by
\begin{eqnarray}\label{x15}
 {n}_s-1=2\eta-6\epsilon\quad\mbox{and}\quad  {\alpha}_s=16\epsilon\eta-24\epsilon^2-2\xi,
\end{eqnarray}
where  the second order slow roll parameter
\begin{eqnarray}
\xi\equiv \left(2\epsilon-\frac{\dot\eta}{{ H}\eta}\right)
\eta\cong \frac{V_{\varphi}V_{\varphi\varphi\varphi}}{V^2},
\end{eqnarray}
has been introduced.

Moreover, in inflationary cosmology, the tensor/scalar ratio, namely $r$,  is related with the slow roll parameter $\epsilon$, via the following consistency relation
${r}=16\epsilon.$

The other important parameter that we will use in this work is the running of the running
$\beta_s=\frac{d \alpha_s}{d \ln k}$, given, in the slow roll approximation, by \cite{Ade,Huang}
\begin{eqnarray} \label{eq:betas}
 \beta_s=-192{\epsilon}^3+192{\epsilon}^2\eta-32\epsilon{\eta}^2-24\epsilon\xi+2\eta\xi+2\zeta,
\end{eqnarray}
where we have introduced the third order slow roll parameter
\begin{eqnarray} \label{eq:zeta}
 \zeta\equiv \left(4\epsilon-\eta-\frac{\dot{\xi}}{H\xi}\right)\xi\cong \frac{V^2_{\varphi}V_{\varphi\varphi\varphi\varphi}}{V^3}.
\end{eqnarray}

\medskip

\section{Analytical fitting of the parameters} \label{s:analytical}

\subsection{PLANCK2013 data: the running} \label{ss:planck2013}
In this section we will take  into account the PLANCK2013 constrains on $n_s$, $\alpha_s$ and ${r}$
which are model dependent (see table $5$ of \cite{Ade}),
and the fact  that more than $50$ e-folds are needed to solve the horizon and flatness problems of GR.
If one does not consider the running and makes the analysis in the plane $(n_s,r)$, then  PLANCK2013 data  shrink the space of allowed
standard inflationary models, preferring potentials with a concave shape ($V_{\varphi\varphi}<0$) \cite{Ade}.
But,
it is the combination of the three data $(n_s,\alpha_s,{r})$ what rules out all the standard slow roll inflationary models \cite{running}.

Effectively, for instance, we will consider in detail the $\Lambda$CDM$+r+\alpha_s$ model from PLANCK2013 combined with WP and BAO data, which gives the following results (see table 5  of $\cite{Ade}$).:
$$n_s=0.9607\pm 0.0063,\quad r\leq 0.25 \quad \mbox{at}\quad 95\% \quad\mbox{C.L.} \quad \mbox{and} \quad\alpha_s=-0.021^{+0.012}_{-0.010}.$$

In slow roll inflation, a simple calculation leads to the relation
\begin{eqnarray}
\alpha_s=\frac{1}{2}(n_s-1)r+\frac{3}{32}r^2-2{\xi}.
\end{eqnarray}

And thus, assuming that the  potentials we consider have spectral index $n_s$ at less than 2$\sigma$ deviations and
satisfy the conservative bound $r\leq 0.32$ (see figure 4 of $\cite{Ade}$),  the minimum of the function
$ \frac{1}{2}(n_s-1)r+\frac{3}{32}r^2$ reached at the point ($n_s=0.9481, r=0.1384$)
is greater than $-0.0018$, what provides the bound
\begin{eqnarray}\label{X}
\alpha_s\geq -0.0018-2\xi,
\end{eqnarray}
meaning that plateau potentials such as $V(\varphi)=V_0\left(1-\frac{\varphi^2}{\mu^2}\right)$
(Hill-Top Inflation (HTP)) \cite{albrecht}, $V(\varphi)=V_0\left(1-\frac{\varphi^2}{\mu^2}\right)^2$ with $|\varphi|\leq \mu$ (Double-Well Inflation (DWI)) \cite{olive} or
$V(\varphi)=V_0\left(1+\cos\left(\frac{\varphi}{\mu}\right)\right)$ (Natural Inflation (NI)) \cite{freese},
when one considers values of the running at 1$\sigma$ C.L.,
are disfavoured
by PLANCK data because for all of them $\xi\leq 0$. In fact, the deviation from the theoretical value of the running to its expected observable value, namely ${\frak D}$, is larger than
$1.6\sigma$.

A distance larger than $1.6\sigma$ is also obtained for
the potential that leads to Exponential SUSY Inflation (ESI) $V(\varphi)=V_0\left(1-e^{-p\varphi}\right)$ \cite{stewart}
and for Power Law Inflation (PLI) whose potential is given by $V(\varphi)=V_0e^{-p\varphi}$ \cite{lucchin}, because in this
case one has $\xi=\eta^2$, that is,
\begin{eqnarray}
\alpha_s=\frac{r}{8}\left((n_s-1)+\frac{3}{16}r \right)-\frac{1}{2}(n_s-1)^2=-\frac{3r^2}{32}\geq -0.0018,
\end{eqnarray}
where we have evaluated $\alpha_s$, as a function of $n_s$ and $r$  at the absolute minimum, namely $n_s=1-\frac{3r}{8}$ with $r=0.1384$, in the rectangle $ [0.9481, 0.9733]\times [0,0.32]$.
To be more precise, since  power law inflation has no running, the distance to the main observational value is $1.7\sigma$, because in this case one has the bound $\alpha_s\geq -0.00045$.

Dealing with the K\"aller Moduli Inflation I (KMII) \cite{conlon} given by the potential $V(\varphi)=V_0\left(1-\alpha\varphi e^{-\varphi}\right)$,
with Higgs Inflation (HI) \cite{Bezrukov} where the potential is
$V(\varphi)=V_0\left(1- e^{-\sqrt{\frac{2}{3}}\varphi}\right)^2$ and for Open String Tachyonic Inflation (OSTI) \cite{Witten1} with potential
$V(\varphi)=-V_0\left(\frac{\varphi}{\mu} \right)^2\ln\left(\frac{\varphi}{\mu} \right)^2$, one always has
$\xi\leq \eta^2$ and one obtains the same conclusions as in  ESI, i.e., ${\frak D}\geq 1.6\sigma$.

 When one considers Large Field Inflation (LFI) given by the monomial  potential $V(\varphi)=V_0\varphi^{p}$ with $p\geq 1$ \cite{Lindea}, one obtains
\begin{eqnarray}\label{zzw}
 n_s-1=-\frac{p(p+2)}{\varphi^2}\quad\mbox{and}\quad  \alpha_s=-\frac{2p^2(p+2)}{\varphi^4},
\end{eqnarray}
which means that $p$ must be positive in order to have an spectral index with a red tilt and a negative running. As a first consequence, Inverse Monomial Inflation (IMI) \cite{barrow} is
disfavored. Second, from the relation
\begin{eqnarray}\alpha_s=-\frac{2}{p+2}(n_s-1)^2\geq -\frac{2}{3}(n_s-1)^2\geq -0.0018,\end{eqnarray}
we conclude that
${\frak D}\geq 1.6\sigma$.
And for
 Radiation Gauge Inflation (RGI) \cite{honorez},  whose potential is given by $V(\varphi)=V_0\frac{\varphi^2}{\alpha+\varphi^2}$ one has
$\xi=\frac{12\varphi^2(\varphi^2-\alpha)}{(3\varphi^2-\alpha)^2}\eta^2\leq \frac{4}{3}\eta^2$, because $\frac{12\varphi^2(\varphi^2-\alpha)}{(3\varphi^2-\alpha)^2} $
increases as a function of $\varphi$. Then, evaluating at $n_s=0.9481$ one has
\begin{eqnarray}
 \alpha_s=-\frac{2}{3}(n_s-1)^2\geq -0.0018,
\end{eqnarray}
giving as a result ${\frak D}\geq 1.6\sigma$.

For potentials such as:
$V(\varphi)=V_0\left(1-\left(\frac{\varphi}{\mu}\right)^{-p}\right)$ (Brane Inflation (BI)) \cite{espinosa},
and $V(\varphi)=V_0\left(1+\left(\frac{\varphi}{\mu}\right)^{-p}\right)$ (Dynamical Supersymmetric Inflation (DSI))  \cite{casas},  since one has
$\xi=\frac{p+2}{p+1}\eta^2$, one can obtain
the following exact formula
\begin{eqnarray}\label{AA}
\alpha_s=\frac{(p-2)r}{8(p+1)}\left((n_s-1)+\frac{3r}{16}\right)
-\frac{p+2}{2(p+1)}(n_s-1)^2.
\end{eqnarray}

The minimum of $\alpha_s$ is obtained at $n_s=1-\frac{3r}{8}$ with $r=0.1384$, and thus, inserting this expression in (\ref{AA}), one gets
\begin{eqnarray}
\alpha_s\geq-\frac{3r^2}{32}\geq -0.0018,
\end{eqnarray}
which is incompatible with the running provided by PLANCK at $1\sigma$ C.L., because
${\frak D}\geq 1.6\sigma$.

For  general hill-top potentials such as: $V(\varphi)=V_0\left(1-\left(\frac{\varphi}{\mu}\right)^p\right)$ (Small Field Inflation (SFI))  \cite{albrecht} and
$V(\varphi)=V_0\left(1+\left(\frac{\varphi}{\mu}\right)^p\right)$
(Valley Hybrid Inflation (VHI))  \cite{LINDE}
with $p\geq 3$,  since one
has
$\xi=\frac{p-2}{p-1}\eta^2$, one can obtain
the following exact formula
\begin{eqnarray}
\alpha_s=\frac{(p+2)r}{8(p-1)}\left((n_s-1)+\frac{3r}{16}\right)
-\frac{(p-2)(n_s-1)^2}{2(p-1)}.
\end{eqnarray}

Since the minimum of $\alpha_s$ is obtained at $n_s=1-\frac{3r}{8}$ with $r=0.1384$, one also has
\begin{eqnarray}
\alpha_s\geq-\frac{3r^2}{32}\geq -0.0018,
\end{eqnarray}
giving ${\frak D}\geq 1.6\sigma$.

However when one deals  with Arctan Inflation (AI) \cite{Wang} with potential $V(\varphi)=V_0\left(1-\frac{2}{\pi}\arctan\left( \frac{\varphi}{\mu} \right)\right)$, where one has
$\xi\leq \frac{3}{2}\eta^2$. The absolute minimum is reached at the point ($n_s=0.9481, r=0.32$), leading to the constrain
\begin{eqnarray}
 \alpha_s=-\frac{(n_s-1)r}{16}-\frac{3r^2}{256}-\frac{3(n_s-1)^2}{4}\geq -0.0022,
\end{eqnarray}
which means that the deviation to the expected observational value is greater than $1.56\sigma$.

In  the case of Loop Inflation (LI) \cite{binetruy} with potential $V(\varphi)=V_0(1+\alpha\ln\varphi)$, one has $\xi=2\eta^2$, leading to the constrain
\begin{eqnarray}
 \alpha_s=-\frac{1}{4}(n_s-1)r-\frac{3}{64}r^2-(n_s-1)^2\geq -0.0034,
\end{eqnarray}
when one evaluates at the point where $\alpha_s$ reaches its absolute minimum, namely ($n_s=0.9481, r=0.32$).
Consequently, ${\frak D}\geq 1.46\sigma$.

For Mixed Large Field Inflation (MLFI) \cite{bellini} with potential $V(\varphi)=V_0\varphi^2\left(1+\alpha\varphi^2\right)$, one has $\xi=\frac{3\alpha}{2(1+2\alpha\varphi^2)}r$. On
the other hand,
the relation
\begin{eqnarray}
r=16\epsilon=\frac{32}{\varphi^2}\left(\frac{1+2\alpha\varphi^2}{1+\alpha\varphi^2} \right)^2\geq \frac{32}{\varphi^2},
\end{eqnarray}
leads to the bound $\frac{1}{1+2\alpha\varphi^2}\leq \frac{r}{r+64\alpha}\leq\frac{r}{64\alpha} $, and thus, $\xi\leq \frac{3r^2}{128}$. Finally, evaluating at
the  absolute minimum ($n_s=0.9481, r=0.276$) we can conclude
\begin{eqnarray}
 \alpha_s\geq\frac{1}{2}(n_s-1)r+\frac{3}{64}r^2=-\frac{3r^2}{64}\geq -0.0036,
\end{eqnarray}
and thus, ${\frak D}\geq 1.45\sigma$.

Finally,  in Witten-O'Raifeartaigh Inflation (WRI) \cite{Witten} with potential
$V(\varphi)=V_0\ln^2\left(\frac{\varphi}{\mu} \right)$ one has $\xi\leq \frac{9}{4}\eta^2$, leading to the constrain
\begin{eqnarray}
 \alpha_s\geq-\frac{11}{32}(n_s-1)r-\frac{33}{512}r^2-\frac{9}{8}(n_s-1)^2\geq -0.0040,
\end{eqnarray}
when one evaluates at the point where $\alpha_s$ reaches its absolute minimum, namely ($n_s=0.9481, r=0.32$).
Consequently, ${\frak D}\geq 1.42\sigma$.

\medskip

To end this section a remark is in order: First of all, it is important to realize that we have studied analytically only 25 of the 49 models provided by \cite{MARTIN}. Secondly,
our analytic results have been obtained bounding the minimum of $\alpha_s$ in the rectangle $$R=\{(n_s,r):0.9481\leq n_s\leq  0.9733,\quad  0\leq r\leq 0.32\}.$$
However, the running could be parametrized with only one independent variable, for instance, the scalar field $\varphi$, which means that the bounds obtained in the rectangle could
be improved, because one only needs to find a bound inside a curve inside the rectangle $R$. The problem to perform this calculation analytically is that this curve can only be obtained
explicitly for a few potentials. The numerical calculations of Section \ref{s:numerical} are needed to improve the analytical ones.


\subsection{Analytical results from PLANCK2013:$\Lambda$CDM+$r$+$\alpha_s$ model combined with   other data}

The same kind of results could be obtained from  other models with running.
In fact, assuming that the potentials we choose have an spectral index $n_s$ at less than $2\sigma$ deviations and
 also they satisfy the conservative
constrain $r\leq 0.32$, we have summarized the results in four  Tables: Table 1  contains the description of three models and Table 2 (WP+high-${{\ell}}$ ), Table 3 (WP) and Table 4 (WP+lensing) contain the deviation of the running, for the potentials we have analytically studied, for each one of the models. Again, the computed likelihoods reflect that the values of the running $\alpha_s$ predicted by the
potentials lie all in the same tail of the Gaussian distribution.

\begin{table}[ht]
\label{t:planck2015nobetas}
\begin{tabular}{|l|c|c|c|}
\hline
Determination & $r$ & $n_s$ & $\alpha_s$ \\
\hline
PLANCK2013+WP+ high-${{\ell}}$ & $\leq 0.23$ & $0.9570\pm0.0075$ & $-0.022^{+0.011}_{-0.010}$ \\
\hline
PLANCK2013 +WP& $\leq 0.25$ & $0.9583 \pm 0.0081$ &
$-0.021\pm 0.012$ \\
\hline
PLANCK2013 +WP+lensing & $\leq 0.26$ & $0.9633\pm0.0072$ & $-0.017\pm 0.012$ \\
\hline
\end{tabular}
\caption{PLANCK2013 estimations of spectral parameters, without running of the running
(\cite{Ade})
.}
\end{table}
  \begin{table}[ht]
\begin{tabular}{|l|c|c|}
\hline
Potential & Running deviation & Disfavored at   \\
\hline
PLI&  $\geq 2 \sigma$& 97.75 \% C.L. or more   \\
\hline HTI, DWI, NI,  ESI, KMII, HI, PSNI & & \\
SFI, LFI, VHI, DSI, BI, OSTI, RGI
 & $\geq 1.8 \sigma$ &   96.4 \% C.L. or more  \\
\hline
AI &$\geq 1.7\sigma$ &   95.55 \% C.L. or more  \\
\hline
 LI, MLFI, WRI &$\geq 1.6\sigma$ &   94.5 \% C.L. or more  \\
\hline
\end{tabular}
\caption{ PLANCK2013+WP+high-${{\ell}}$:$\Lambda$CDM+$r$+$\alpha_s$ model.}
\end{table}
 \begin{table}[ht]
\begin{tabular}{|l|c|c|}
\hline
Potential & Running deviation & Disfavored at   \\
\hline
PLI&  $\geq 1.7 \sigma$& 95.55  \% C.L. or more   \\
\hline HTI, DWI, NI,  ESI, KMII, HI, PSNI & & \\
SFI, LFI, VHI, DSI, BI, OSTI, RGI,  AI
 & $\geq 1.5 \sigma$ &   93.93 \% C.L. or more  \\
\hline
LI &$\geq 1.4\sigma$ &   91.95 \% C.L. or more  \\
\hline
  MLFI, WRI &$\geq 1.3 \sigma$ &   90.3 \% C.L. or more  \\
\hline
\end{tabular}
\caption{PLANCK2013+WP:$\Lambda$CDM+$r$+$\alpha_s$ model.}
\end{table}
 \begin{table}[ht]
\begin{tabular}{|l|c|c|}
\hline
Potential & Running deviation & Disfavored at   \\
\hline
PLI&  $\geq 1.4 \sigma$& 91.95 \% C.L. or more  \\
\hline HTI, DWI, NI,  ESI, KMII, HI, PSNI & & \\
SFI, LFI, VHI, DSI, BI, OSTI, RGI,  AI
 & $\geq 1.2 \sigma$ &   88.5 \% C.L.  or more \\
\hline
 LI, MLFI, WRI &$\geq 1.1 \sigma$ &   86.45 \% C.L. or more  \\
\hline
\end{tabular}
\caption{For PLANCK2013+WP+lensing:$\Lambda$CDM+$r$+$\alpha_s$ model.}
\end{table}

\newpage

\subsection{PLANCK2015 data: the running almost vanishes}

The new observational data provided by \cite{Planck:2015xua}, reproduced
in Table 5
reduces one order or more (depending on other data such as BAO, lensing and lowP),
the modulus of the expected observational value of the running.

\begin{table}[ht]
\label{t:planck2015nobetas}
\begin{tabular}{|l|c|c|c|}
\hline
Determination & $r$ & $n_s$ & $\alpha_s$ \\
\hline
PLANCK2015 TT+ lowP & $\leq 0.180$ & $0.9667\pm0.0066$ & $-0.0126^{+0.0098}_{-0.0087}$ \\
\hline
PLANCK2015 TT+ lowP+lensing & $\leq 0.186$ & $0.9690 \pm 0.0063$ &
$-0.0076^{+0.0092}_{-0.0080}$ \\
\hline
PLANCK2015 TT+ lowP+BAO & $\leq 0.176$ & $0.9673\pm0.0043$ & $-0.0125\pm 0.0091$ \\
\hline
PLANCK2015 TT, TE, EE+ lowP & $\leq 0.152$ & $0.9644\pm0.0049$ & $-0.0085\pm 0.0076$ \\
\hline
\end{tabular}
\caption{PLANCK2015 estimations of spectral parameters, without running of the running
(\cite{Planck:2015xua}).}
\end{table}

 This drastic reduction of the running in modulus is what allows the viability of the potentials disregarded from PLANCK2013 data. Effectively,
 using for instance, the PLANCK2015 TT+ lowP+BAO:$\Lambda$CDM+$r$+$\alpha_s$ data,
 and assuming that the potentials have spectral index  $n_s$ at less than $2\sigma$ deviations,  and also
 satisfying the conservative bound $r\leq 0.32$. Then,  for the simplest potentials, analytically  one can show:
  \begin{enumerate}
  \item For LFI, one gets the following bound
  \begin{eqnarray}
   |\alpha_s|\leq \frac{2}{2+p}(n_s-1)^2\leq (n_s-1)^2\leq 0.0017,
\end{eqnarray}
which means ${\frak D}\leq 1.2\sigma$, and thus, disfavoring the potential less than the 86.45 \% C.L..

\item
For  PLI, which has no-running, one obtains ${\frak D}\leq 1.4\sigma$, and thus, disfavoring the potential less than the 91.95 \% C.L..

\item
For HTP, ESI, BI, DSI, SFI and VHI the minimum and maximum of $\alpha_s$ are obtained respectively at $n_s=1-\frac{3r}{8}$
with $r=0.1101$ and $r=0.0642$. And thus, one gets $-0.0012\leq \alpha_s\leq -0.0003$, what implies
${\frak D}\leq 1.4\sigma$, and thus, disfavoring the potential less than the 91.95 \% C.L..
\end{enumerate}

The rest of potentials in the examined list fit even better the values of the spectral parameters
of Table 5
for some choice of parameter or field values.

\medskip

\subsection{PLANCK2015 data: the running of the running}
The last PLANCK2015 data about the running $\alpha_s$ and its  running $\beta_s=
\frac{d \alpha_s}{d \ln k}$ are reproduced in Table 6.

\begin{table}[ht]
\label{t:planck2015betas}
\begin{tabular}{|l|c|c|c|}
\hline
Determination & $n_s$ & $\alpha_s$ & $\beta_s$ \\
\hline
PLANCK2015 TT+lowP & $0.9569\pm 0.0077$ & $0.011^{+0.014}_{-0.013}$ &
$0.029^{+0.015}_{-0.016}$ \\
\hline
PLANCK2015 TT,TE,EE+lowP & $0.9586\pm 0.0056$ & $0.009\pm 0.010$ & $0.025\pm 0.013$ \\
\hline
\end{tabular}
\caption{Determinations of the spectral parameter values, with running of the running by PLANCK2015 ((19) of \cite{Planck:2015xua})}
\end{table}

These values are similar to the ones of PLANCK2013, reproduced in Table 7.

\begin{table}[ht]
\label{t:planck2013betas}
\begin{tabular}{|l|c|c|c|}
\hline
Determination & $n_s$ & $\alpha_s$ & $\beta_s$ \\
\hline
PLANCK2013+WP+BAO & $0.9568^{+0.068}_{-0.063}$ &
$0.000^{+0.013}_{-0.016}$ & $0.017^{+0.016}_{-0.014}$ \\
\hline
PLANCK2013+WP+high-$\ell$ &
$0.9476^{+0.086}_{-0.088}$ & $0.001^{+0.013}_{-0.014}$ & $0.022^{+0.016}_{-0.013}$ \\
\hline
PLANCK2013+WP+lensing & $0.9573^{+0.077}_{-0.079}$ & $0.006^{+0.015}_{-0.014}$ &
$0.019^{+0.018}_{-0.014}$ \\
\hline
PLANCK2013+WP & $0.9514^{+0.087}_{-0.090}$ & $0.001^{+0.016}_{-0.014}$ &
$0.020^{+0.016}_{-0.015}$ \\
\hline
\end{tabular}
\caption{Determinations of the spectral parameter values, model $\Lambda$CDM+$\alpha_s$+$\beta_s$, by PLANCK2013 (table 5 of \cite{Ade})}
\end{table}

These results contradict single field slow roll inflation, because in that case, the running $\alpha_s$ is second order in the slow roll parameters, and its running is given by Eqs.
\eqref{eq:betas}, \eqref{eq:zeta},
which make $\beta_s$ a third order parameter, while the values determined by
PLANCK place $\beta_s$ in a higher order of magnitude than the running $\alpha_s$
itself. Moreover, disregarding the running of the running, the running is negative while taking into account it, the running becomes positive. This seems a signature of
the problem that suffers the method used to reconstruct the power spectrum of scalar perturbations from observational data:
the value of the coefficients in the Taylor series of the power spectrum logarithm function could suffer a bias. We will address this question in Section 5.

To show the improbability of the observed value of $\beta_s$
analytically for all the potentials that appear in \cite{MARTIN}
is very involved due to the increasing complexity of the formulas \eqref{eq:betas}, \eqref{eq:zeta} for the new parameter $\beta_s$. This work
thus follows a double approach, combining analytical
calculations  for  LFI, LI (with $\alpha>0$), VHI, SFI, BI and  DSI (with $p$ and even number or $\varphi\geq 0$),
  HTI, ESI and  PLI,  with a numerical
analysis to which we subject all the potentials in \cite{MARTIN}.

Let us start with the analytical approach. In the  case of LFI one has
\begin{eqnarray}
 n_s-1=-\frac{p(p+2)}{\varphi^2},\quad \alpha_s=-\frac{2p^2(p+2)}{\varphi^4}\quad\mbox{and}\quad \beta_s=-\frac{8p^3(p+2)}{\varphi^6},
\end{eqnarray}
and thus,
\begin{eqnarray}\label{XXX}
 \beta_s=-\frac{8}{(p+2)^2}(n_s-1)^3\Longrightarrow |\beta_s|\leq \frac{8}{9}|n_s-1|^3.
\end{eqnarray}

Then, for the PLANCK2015 TT,TE,EE+lowP data, considering $n_s$ at $2\sigma$ C.L., and thus,  after inserting $n_s=0.9474$ in (\ref{XXX}), one obtains the bound $\beta_s\leq 0.00013$,
which means that the deviation from the theoretical value of the running of the running to its expected observational value is larger than $1.9\sigma$. A deviation larger than
$1.9\sigma$ is also obtained from  PLANCK2015 TT+lowP data, meaning that LFI is completely disfavored by PLANCK2015 data.

In the case of  LI (with $\alpha>0$), VHI, SFI, BI and  DSI (with $p$ and even number or $\varphi\geq 0$), HTI, ESI and  PLI,
a simple calculation shows that $\beta_s\leq 0$ which means that the deviation from the theoretical value of the running of
the running to its expected observational value  is also larger than $1.9\sigma$.
\medskip

\section{Numerical fitting of the parameters} \label{s:numerical}

\subsection{Methodology} \label{ss:algoritme}

Let us describe the numerical tests that the authors have applied to all single field inflationary models from the list of \cite{MARTIN}. These tests have been built into a MATLAB program that takes as input
a list of potentials $V(\varphi)$ and values for spectral parameters in the list $r,n_s,\alpha_s,\beta_s$,
and asseses the likelihood of each model in the list to fit the values of the spectral parameters using a 95\% confidence limit for the value of $r$, and assuming Gaussian
distribution for the values of $n_s, \alpha_s, \beta_s$.

For each cosmological model, a broad
range of possible values for the parameters on which it depends has been determined
following \cite{MARTIN}.
A test list of values for each parameter has been selected, covering in a
dense, equispaced fashion finite intervals of possible values for the parameter,
and approaching with log-equispaced values every finite or infinite limit value
for the parameter.

After allowing simplifications induced by rescaling,
if a model still depended on more than one parameter all possible combinations
of values for each parameter were tested. Table \ref{t:llistamodels} lists the selected
parameter values for each model.

Also separately for each model, the range of possible values for the inflaton
field was determined taking into account whether the model admitted values of the field with any sign, or only positive values, and further peculiarities such
as periodicity
of the model and the applied rescalings. The considered range $[\varphi_0,\varphi_f]$
for field values in each model is also listed in Table \ref{t:llistamodels}.

The MATLAB software developed by the authors,
for each model $V(\varphi)$ and choice of value of the parameter(s) on which it depends,
takes an equispaced mesh of values in the range $[\varphi_0,\varphi_f]$ of possible values of the field in this model. This mesh is taken increasingly fine, currently up to
step $\Delta \varphi = 2 \cdot 10^{-4}$.

The subintervals in the range of field values for which the potential satisfies
$V(\varphi)>0$ are numerically determined over the selected mesh, and each
interval of positive values of the potential for the selected values
of the model parameters is considered as a {\em case}, which thus consists of:
\begin{itemize}
\item a candidate theory with a given potential $V(\varphi)$,
\item a specific choice of
parameter values for $V$,
\item and a range of values $[\bar \varphi_0,\bar \varphi_f]$ of the inflaton field $\varphi$
such that $V>0$ on them.
\end{itemize}

The numerical test for each case consists in meshing the interval of field
values with a uniform step (of size $\Delta \varphi = 2 \cdot 10^{-4}$ for the results reported
in this work), computing the spectral parameters $r, n_s, \alpha_s, \beta_s$
for each value of the field $\varphi$ in the mesh using the formulas
of Section \ref{s:formules} and symbolic derivation of the potential $V$
to produce the derivatives $V_\varphi, \dots, V_{\varphi \varphi \varphi \varphi}$,
 and then applying successive filtering criteria to determine
which values of the field $\varphi$ fulfill simultaneously all of them, thus allowing the
model in this particular case to fit the spectral measured data for which the model is tested.

\medskip

A determination of a set of values for the spectral parameters provides
their expected values $<n_s>,<\alpha_s>,<\beta_s>$, and their standard
deviations $\sigma_{n_s},\sigma_{\alpha_s},\sigma_{\beta_s}$. In the case
of spectral parameters without running of the running $\beta_s$, this
is replaced by the tensor/scalar ratio $r$, which is conservatively estimated
to have a value $r<0.32$ (see \cite{Ade},\cite{Planck:2015xua}).

\medskip

The applied filters in the case of spectral parameters without running
of the running consist in looking for the values of the field $\varphi$ such that:
\begin{enumerate}
\item $r<0.32$,
\item $|n_s(\varphi)-<n_s>| < 2\sigma_{n_s}$,
\item $|\alpha_s(\varphi)-<\alpha_s>| < 1.6\sigma_{\alpha_s}$,
\end{enumerate}
A model not passing the first filter is ruled out with 95\% C.L.,
a model not passing the second filter is ruled out with 95.5\% C.L.,
and a model not passing the third filter is ruled out with 94.5\% C.L.,
because the values of $\alpha_s$ provided by the models are located
uniformly in the same tail of the Gaussian distribution (namely, from the
negative expected value towards zero) for the spectral valuations
to which we have applied our test.

\medskip

The applied filters in the case of spectral parameters with running
of the running consist in looking for the values of the field $\varphi$ such that:
\begin{enumerate}
\item $\epsilon(\varphi) \le 1$,
\item $|n_s(\varphi)-<n_s>| < 2\sigma_{n_s}$,
\item $|\alpha_s(\varphi)-<\alpha_s>| < 2\sigma_{\alpha_s}$,
\item $|\beta_s(\varphi)-<\beta_s>| < 1.8\sigma_{\beta_s}$,
\end{enumerate}
A model not passing the first filter (equivalent to asking for $r<16$)
is ruled out with 95\% C.L.,
a model not passing the second or third filter is ruled out with 95.5\% C.L.,
and a model not passing the fourth filter is ruled out with 92.8\% C.L.

\medskip

In either situation with or without running of the running,
a case (choice of model, parameter values, and interval of values for
the field) is considered possible only if there exists some value of the field $\varphi$
in its range such that it satisfies simultaneously all of the filtering conditions.
The filters are not independent, thus if a case does not satisfy
all of the filtering criteria simultaneously for any field value $\varphi$ in its
range, it is disproved with the confidence level of
the strongest filter it fails. A model such that it is disproved
in any case (i.e., for any choice of parameters and range of field values)
is regarded as disproved with the lowest confidence level with which any
of its cases is disproved.

Conversely, if a case has values of the field $\varphi$ satisfying all of the filtering
criteria, the minimal values among them of the distances $n_s(\varphi)-<n_s>,
\alpha_s(\varphi)-<\alpha_s>,\beta_s(\varphi)-<\beta_s>$, expressed in standard deviations, are logged. The maximum among these distances provides the confidence level to which
the particular case has been disproved. In the case of computations without
running of the running such that all of the values of $\alpha_s$ provided by
the models are found consistently in the same tail of the Gaussian distribution
their likelyhood is assessed taking into account only this tail.

In these cases satisfying all filters, i.e. that are not
disproved with C.L. at least $92.8\%$ the testing software also looks for
values of the field $\varphi_e$ such that $\epsilon(\varphi_e) \cong 1$,
and using them as endpoints of the inflationary phase,
computes the number of e-folds of inflation for any choice of $\varphi$
in the case, by integrating numerically with a trapezoidal rule
\begin{eqnarray}
N(\varphi)=\left|\int_{\varphi_e}^{\varphi} \frac{V}{V_\varphi} d\varphi\right|
\end{eqnarray}
The result $N(\varphi)$ ranges over the number of e-folds of expansion
that the case supports for the field values $\varphi$ satisfying all
of the filtering criteria. Its minimal and maximal values are logged, as
they are the limit values for the number of e-folds of expansion
that the case can support.

\subsection{Numerical results}

Single-field inflaton models were exhaustively studied in \cite{MARTIN},
from which we take the list of models and parameters to be numerically tested.
Table \ref{t:llistamodels}, adapted from Table 1 of \cite{MARTIN},
presents each model's potential, the range of values of the parameters for which
it has been tested, and the range of values of the inflaton field over
which it has been tested.

{\tiny
\begin{longtable}{|c|c|c|c|}
\caption{
Models from \cite{MARTIN} numerically tested.
}
\label{t:llistamodels}
\\
\hline
  Name  & $V(\varphi)$ & Parameter values & Field values \\
  \hline \hline
  HI & $V_0\left(1-e^{-\sqrt{2/3}\varphi}\right)^2$ &  & [-40,40] \\
  \hline \hline
  RCHI & $V_0\left(1-2e^{-\sqrt{2/3}\varphi}+\frac{A_I}{16\pi^2}
 \frac{\varphi}{\sqrt{6}}\right)$ & $A_I$: [linspace(-100,100,120),linspace(-3,3,200)] & [-10,20] \\
  \hline \hline
  LFI & $V_0\left(\varphi\right)^p$ & $p$: linspace(0.5,20,60) &  \\
  \hline \hline
  MLFI & $V_0\varphi^2 \left[1 + \alpha \varphi^2 \right]$ & $\alpha$: [linspace(-10,100,61),linspace(-0.1,0.1,120)] &  \\
  \hline \hline
  RCMI & $V_0\left(\varphi\right)^2
\left[1-2\alpha\varphi^2\ln \left(\varphi\right)\right]$ & $\alpha$: [linspace(1e-4,1.5,30),10.$\wedge$linspace(-14,-5,20)] &  \\
  \hline \hline
  RCQI & $V_0\left(\varphi\right)^4
\left[1-\alpha \ln\left(\varphi \right)\right]$ & $\alpha$: [linspace(1e-2,10,120),10.$\wedge$linspace(-6,-2.5,10)] &  \\
  \hline \hline
  NI & $V_0\left[1+\cos\left(\frac{\varphi}{f}\right)\right]$ & $f$: 1 & $[10^{-5},\pi]$ \\
  \hline \hline
  ESI & $V_0\left(1-e^{-q\varphi}\right)$ & $q$: [linspace(0.1,10,120),linspace(1e-5,0.099,60)] &  \\
  \hline \hline
  PLI & $V_0e^{-\alpha \varphi}$ & $\alpha$: [linspace(0.1,10,120),10.$\wedge$linspace(-6,-1.5,15)] & [-40,40] \\
  \hline \hline
  KMII & $V_0\left(1-\alpha\varphi e^{-\varphi}\right)$ & $\alpha$: [linspace(1e-2,10,60),10.$\wedge$linspace(-6,-2.5,10)] &  \\
  \hline \hline
  HF1I & $V_0 \left(1+A_1 \varphi\right)^2\left[1-\frac{2}{3}
\left(\frac{A_1}{1+A_1\varphi}\right)^2\right]$ & $A_1$: linspace(1e-3,40,180) & [-40,40] \\
  \hline \hline
  CWI & $V_0\left[1 +
\alpha\left(\frac{\varphi}{Q}\right)^4 \ln
\left(\frac{\varphi}{Q}\right)\right]$ & $Q$: [10.$\wedge$linspace(-6,-2.5,12),linspace(1e-2,10,60)] &  \\
  \hline \hline
  LI & $V_0\left[1
+\alpha\ln \left(\varphi\right)\right]$ & $\alpha$: [linspace(-0.3,0.3,60),-10.$\wedge$linspace(-1,0,10),10.$\wedge$linspace(-1,0,10)] &  \\
  \hline \hline
  RpI & $V_0 e^{-2 \sqrt{2/3}\varphi} \left|e^{\sqrt{2/3}\varphi}
 - 1 \right|^{2p/(2p-1)}$  & $p$: linspace(0.25,10,60) &  \\
  \hline \hline
  DWI & $V_0\left[\left(\frac{\varphi}{\varphi_0}\right)^2-1\right]^2$ & $\varphi_0$: 1 & $[10^{-4},80]$ \\
  \hline \hline
  MHI & $V_0 \left[1-\text{sech} \left(\frac{\varphi}{\mu} \right) \right]$ & $\mu$: 10 & $[10^{-4},400]$ \\
  \hline \hline
  RGI & $V_0\frac{\left(\varphi\right)^2}{\alpha+\left(\varphi\right)^2}$ & $\alpha$: [linspace(1e-1,10,30),10.$\wedge$linspace(-6,-1.5,12)] &  \\
  \hline \hline
  MSSMI & $V_0\left[\left(\frac{\varphi}{\varphi_0}\right)^2-\frac{2}{3}
  \left(\frac{\varphi}{\varphi_0}\right)^6+\frac{1}{5}\left(
  \frac{\varphi}{\varphi_0}\right)^{10}\right]$ & $\varphi_0$: [1e-7,1e-3,1] & $[10^{-4},80]$ \\
  \hline \hline
  RIPI & $V_0 \left[ \left(\frac{\varphi}{\varphi_0}\right)^2 -
   \frac{4}{3} \left( \frac{\varphi}{\varphi_0} \right)^3 + \frac{1}{2}
   \left( \frac{\varphi}{\varphi_0} \right)^4 \right]$ & $\varphi_0$: [1e-7,1e-3,1] & $[10^{-4},80]$ \\
  \hline \hline
  AI & $V_0\left[1-\frac{2}{\pi}
    \arctan\left(\frac{\varphi}{\mu}\right)\right]$ & $\mu$: [1e-2,1] & [-40,40] \\
  \hline \hline
  CNAI & $V_0\left[3-\left(3+\alpha^2 \right) \tanh^2
    \left( \frac{\alpha}{\sqrt{2}} \varphi \right) \right]$ & $\alpha$: [linspace(1e-2,20,120),10.$\wedge$linspace(-6,-2.5,10)] &  \\
  \hline \hline
  CNBI & $V_0\left[\left(3-\alpha^2\right) \tan^2
    \left(\frac{\alpha}{\sqrt{2}}\varphi \right)-3\right]$ & $\alpha$: [linspace(1e-3,5,40),10.$\wedge$linspace(-7,-3.5,12)] &  \\
  \hline \hline
  OSTI & $-V_0\left(\frac{\varphi}{\varphi_0}\right)^2\ln\left[\left(\frac{\varphi}{\varphi_0}\right)^2\right]$ & $\varphi_0$: 1 & $[10^{-6},1]$ \\
    \hline \hline
  WRI & $V_0\ln^2 \left(\frac{\varphi}{\varphi_0}\right)$ & $\varphi_0$: 1 & $[10^{-4},10]$ \\
   \hline \hline
  SFI & $V_0 \left[1 -\left(\frac{\varphi}{\mu}\right)^{p}\right]$ & \begin{tabular}{l} $\mu$: 1 \\ $p$: linspace(0.5,10,20) \end{tabular} & $[10^{-4},1]$ \\
  \hline \hline
  II & $V_0\left(\varphi-\varphi_0\right)^{-\beta}
-V_0\frac{\beta ^2}{6}\left(\varphi-\varphi_0\right)^{-\beta-2}$ & \begin{tabular}{l} $\varphi_0$: 0 \\ $\beta$: [linspace(0.1,10,31),linspace(20,50,3)] \end{tabular} &  \\
  \hline \hline
  KMIII & $V_0\left[1-\alpha(\varphi)^{\frac43}\exp\left(-\beta(\varphi)^{\frac43}\right)\right]$ & \begin{tabular}{l} $\alpha$: 10.$\wedge$linspace(-3,12,46) \\ $\beta$: 10.$\wedge$linspace(-3,12,46) \end{tabular} & $[10^{-4},10]$ \\
  \hline \hline
  LMI & $V_0\left(\varphi\right)^{\alpha}\exp\left[-\beta (\varphi)^{\gamma}\right]$ & \begin{tabular}{l} $\beta$: [linspace(0.1,20,40),10.$\wedge$linspace(-4,-1.5,10)] \\ $\gamma$: [linspace(1e-3,2,30),10.$\wedge$linspace(0.5,2,4),10.$\wedge$linspace(-6,-4,3)] \end{tabular} &  \\
  \hline \hline
TWI & $V_0\left[1-A\left(\frac{\varphi}{\varphi_0}
    \right)^2e^{-\varphi/\varphi_0} \right]$ & \begin{tabular}{l} $\varphi_0$: 1 \\ $A$: linspace(0.001,8,120) \end{tabular} &  \\
  \hline \hline
GMSSMI & $V_0\left[\left(\frac{\varphi}{\varphi_0}\right)^2-\frac{2}
{3}\alpha\left(\frac{\varphi}{\varphi_0}\right)^6+\frac{\alpha}
{5}\left(\frac{\varphi}{\varphi_0}\right)^{10}\right]$ & \begin{tabular}{l} $\varphi_0$: 10.$\wedge$linspace(-2,0,3) \\ $\alpha$: [linspace(1e-2,2.5,120),10.$\wedge$linspace(0.5,1.5,3),10.$\wedge$linspace(-4,-2.5,6)] \end{tabular}  &  \\
  \hline \hline
GRIPI & $V_0\left[\left(\frac{\varphi}{\varphi_0}\right)^2-\frac{4}
{3}\alpha\left(\frac{\varphi}{\varphi_0}\right)^3+\frac{\alpha}
{2}\left(\frac{\varphi}{\varphi_0}\right)^{4}\right]$ & \begin{tabular}{l} $\varphi_0$: 10.$\wedge$linspace(-2,0,3) \\ $\alpha$: linspace(0.1,10,120) \end{tabular} & $[10^{-4},20]$ \\
  \hline \hline
BSUSYBI & $V_0\left(e^{\sqrt{6}\varphi} + e^{\sqrt{6} \gamma
  \varphi} \right)$ & $\gamma$: linspace(1e-5,2,200) & [-40,40] \\
  \hline \hline
TI & $V_0\left(1+\cos\frac{\varphi}{\mu}+\alpha\sin^2\frac{\varphi}{\mu}\right)$ & \begin{tabular}{l} $\mu$: 1 \\ $\alpha$: [linspace(0.01,3,80),10.$\wedge$linspace(-4,-2.5,6),10.$\wedge$linspace(1,2,4)] \end{tabular} &  \\
  \hline \hline
BEI & $V_0\exp_{1-\beta}\left(-\lambda \varphi\right)$ & \begin{tabular}{l} $\beta$: [linspace(-5,5,60),10.$\wedge$linspace(1,2,3),10.$\wedge$linspace(-4,-2,3), \\ -10.$\wedge$linspace(1,2,3),-10.$\wedge$linspace(-4,-2,3)] \\ $\lambda$: 1 \end{tabular} & [-100,100] \\
  \hline \hline
PSNI & $V_0\left[1+\alpha \ln \left(\cos\frac{\varphi}{f}\right)\right]$ & \begin{tabular}{l} $\alpha$: [linspace(0.1,10,60),10.$\wedge$linspace(-4,-1.5,12)] \\ $f$: 1 \end{tabular} & $[10^{-4},\pi/2-10^{-4}]$ \\
  \hline \hline
NCKI & $V_0\left[1+\alpha \ln \left(\varphi\right) + \beta
    \left(\varphi\right)^2\right]$ & \begin{tabular}{l} $\alpha$: 10.$\wedge$linspace(-7,0,16) \\ $\beta$: linspace(-10,10,80) \end{tabular} &  \\
  \hline \hline
CSI & $\frac{V_0}{\left( 1-\alpha\varphi \right)^2}$ & $\alpha$: [linspace(0.1,5,100),10.$\wedge$linspace(-4,-1.5,8),10.$\wedge$linspace(1,2,4)] & [-40,40] \\
  \hline \hline
OI & $V_0 \left(\frac{\varphi}{\varphi_0} \right)^{4}\left[
 \left(\ln \frac{\varphi}{\varphi_0} \right)^2- \alpha \right]$ & \begin{tabular}{l} $\varphi_0$: 1 \\ $\alpha$: [10.$\wedge$linspace(-7,-2,18),linspace(0.03,1,40)] \end{tabular} &  \\
  \hline \hline
CNCI & $V_0\left[ \left( 3+\alpha^2 \right) \coth^2
    \left(\frac{\alpha}{\sqrt{2}}\varphi\right)- 3 \right]$ & $\alpha$: [linspace(0.1,5,40),10.$\wedge$linspace(-7,-1.5,12),10.$\wedge$linspace(1,3,5)] &  \\
  \hline \hline
SBI & $V_0\left\lbrace 1 + \left[ -\alpha + \beta\ln
  \left( \varphi \right) \right] \left( \varphi
\right)^4 \right \rbrace$ & \begin{tabular}{l} $\alpha$: 10.$\wedge$linspace(-8,0,27) \\ $\beta$: 10.$\wedge$linspace(-8,0,27) \end{tabular} &  \\
  \hline \hline
SSBI & $V_0\left[1 + \alpha\left(\varphi\right)^2
  + \beta\left( \varphi \right)^4 \right]$ & \begin{tabular}{l} $\alpha$: [-10.$\wedge$linspace(-5,2,24),10.$\wedge$linspace(-5,2,24)] \\ $\beta$: [-10.$\wedge$linspace(-5,2,24),10.$\wedge$linspace(-5,2,24)] \end{tabular} & $[10^{-4},20]$ \\
  \hline \hline
IMI & $V_0\left(\varphi\right)^{-p}$ & $p$: linspace(0.5,10,40) &  \\
  \hline \hline
BI & $V_0 \left[1 -\left(\frac{\varphi}{\mu}\right)^{-p}\right]$  & \begin{tabular}{l} $p$: [linspace(1,10,37),10.$\wedge$linspace(-1,-0.33,3)] \\ $\mu$: [1e-4,1] \end{tabular} &  \\
  \hline \hline
RMI & $V_0\left[1-\frac{c}{2}\left(-\frac{1}{2} +\ln
\frac{\varphi }{\varphi_0}\right)\varphi ^2\right]$ & \begin{tabular}{l} $\varphi_0$: 1 \\ $c$: [-linspace(2,10,33),-10.$\wedge$linspace(-5,0,15),10.$\wedge$linspace(-5,0,14), \\ linspace(2,10,33)] \end{tabular} & $[10^{-4},10]$ \\
  \hline \hline
VHI & $V_0\left[1 +\left(\frac{\varphi}{\mu} \right)^{p} \right]$ & \begin{tabular}{l} $\mu$: 1 \\ $p$: linspace(0.1,12,80) \end{tabular} &  \\
  \hline \hline
DSI & $V_0\left[ 1+\left(\frac{\varphi}{\mu} \right)^{-p} \right]$ & \begin{tabular}{l} $\mu$: 1 \\ $p$: linspace(0.1,12,80) \end{tabular} &  \\
  \hline \hline
GMLFI & $V_0\left(\varphi\right)^p \left[1 + \alpha\left(
  \varphi \right)^q \right]$ & \begin{tabular}{l} $\alpha$: 10.$\wedge$linspace(-7,3,31) \\ $p$: linspace(0.5,12,24) \\ $q$: linspace(0.5,12,24) \end{tabular} & [-40,40] \\
  \hline \hline
LPI & $V_0\left(\frac{\varphi}{\varphi_0}\right)^{p}
  \left(\ln \frac{\varphi}{\varphi_0}\right)^q$ & \begin{tabular}{l} $\varphi_0$: 1 \\ $p$: [linspace(0.5,12,24),10.$\wedge$linspace(1.5,2,2)] \\ $q$: [linspace(0.5,12,24),10.$\wedge$linspace(1.5,2,2)] \end{tabular} &  \\
  \hline \hline
CNDI & $\frac{V_0}{ \left\lbrace 1 + \beta\cos\left[
    \alpha \left( \varphi-\varphi_0 \right) \right] \right \rbrace^2}$
 & \begin{tabular}{l} $\varphi_0$: 0 \\ $\alpha$: [linspace(0.1,1,30),10.$\wedge$linspace(-3,-1.5,4),10.$\wedge$linspace(0.5,2,4)] \\ $\beta$: [linspace(1,10,30),10.$\wedge$linspace(-2,-0.5,8),10.$\wedge$linspace(1.5,2,3), \\ -10.$\wedge$linspace(-2,1,7)] \end{tabular} &  \\  \hline
\caption{\footnotesize
Parameter values expressed in Matlab code: {\tt linspace(a,b,n)} means $n$ equispaced values between $a$ and $b$;
{\tt 10.$\wedge$linspace(a,b,n)} means $n$ log-equispaced values between $10^a$ and $10^b$.
Parameter $V_0$ and the reduced Planck mass $M_{\text{Pl}}$ always set to 1.
The range of studied field values is $\varphi \in [10^{-4},40]$ unless otherwise indicated.}
\end{longtable}}

The models, choice of parameter values and range of field values
 of Table \ref{t:llistamodels} have been subjected to the numerical
test described in subsection \ref{ss:algoritme} for the several determinations
of the spectral parameters $n_s,\alpha_s,\beta_s$ listed in Section \ref{s:analytical}. Let us sum up the
conclusions of the most relevant cases:

\medskip

For the determination of spectral parameters PLANCK2013 combined with WP
and BAO data using the $\Lambda CDM+r+\alpha_s$ model
(i.e. without running of the running) depicted in Table 1, the computation
has concluded that all the models in Table \ref{t:llistamodels}
are disproved within a confidence limit 93\%. The reason in all cases is that
the parameter values that match the bound for $r$ and the expected value of $n_s$
within 2 standard deviations can only provide values of the running $\alpha_s$
that are too close to 0, far above the value
$\alpha_s=-0.021^{+0.012}_{-0.010}$ of this determination, and
consistently in the same tail of the Gaussian distribution.

\medskip

For the determination of spectral parameters PLANCK2015 TT+lowP,
without running of the running, of Table 5
 the result changes
drastically: the models
\begin{enumerate}
\item Natural Inflation (NI) \cite{freese}
\item Power Law Inflation (PLI) \cite{lucchin}
\item Constant $n_s$ A Inflation (CNAI) \cite{Vallinoto}
\item Constant $n_s$ B Inflation (CNBI)  \cite{Vallinoto}
\item Open String Tachyonic Inflation (OSTI) \cite{Witten}
\item Generalised MSSM Inflation (GMSSMI) \cite{Lyth}
\item Generalised Regularised Point Inflation (GRIPI) \cite{Mazumdar}
\item Constant Spectrum Inflation (CSI) \cite{Hodges}
\item Constant $n_s$ C Inflation (CNCI) \cite{Vallinoto}
\item Dynamical Supersymmetric Inflation (DSI) \cite{Kinney},
\end{enumerate}
are disproved with confidence at least 94.5\%,
and asking for the number $N$ of e-folds of expansion that the model
allows to be in the range $[30,80]$ only rules out completely a further
model, namely  Minimal Super-symmetric Standard Model Inflation (MSSMI) \cite{Garcia}. The rest of the models in the table admit
some choice(s) of parameter and field values that
simultaneously satisfies all filtering conditions.

\medskip

The inclusion of the running of the running in the numerical test
results in another reversion of conclusions. For the determination
of spectral parameters PLANCK2015 TT+lowP with running of the running
of Table 6,
the only models in Table \ref{t:llistamodels} which
are not disproved for any choice of parameter and field values
with confidence at least 92.8\% are:


\begin{enumerate}
\item K\"aller Moduli Inflation I (KMII) \cite{conlon}.
\item K\"aller Moduli Inflation II (KMIII) \cite{conlon}.
\item Logamediate Inflation (LMI) \cite{nunes}.
\item Twisted Inflation (TWI) \cite{Davis}.
\item Brane SUSY Breaking Inflation (BSUSYBI) \cite{Dudas}.
\item Spontaneous Symmetry Breaking Inflation (SSBI) \cite{AB}.
\item Running-mass Inflation (RMI) \cite{Covi}.
\item Generalised Mixed Large Field Inflation (GMLFI) \cite{Kinney}.
\item Constant $n_s$ D Inflation (CNDI) \cite{Hodges}.
\end{enumerate}

\noindent (albeit GMLFI is disproved with 92.1\% confidence,
and CNDI with 91.9\% confidence). The most common reason for
disproving the models is now that they only provide values of $\beta_s$
at more than 1.8 standard deviations of distance of the
value 0.029 of this determination, which we point out that
is a quite large value compared with those of $n_s-1=-0.043$
and $\alpha_s=0.011$ in this determination. We will bring up
this subject in the next section.

Of the models that are not disproved for this PLANCK2015 TT+lowP with
running of the running determination of the spectral parameters,
the e-fold test in our software has found only for the models
KMIII,SBI,SSBI,GMLFI,CNDI choices of parameter and field values
passing the filters and further allowing the number of expansion
e-folds $N$ to lie in the range $[30,80]$.

\section{Accuracy and reliability of the spectral parameter values}

The results of the analytical and numerical fitting of the studied
single field inflationary models to the successive sets of values provided by the
Planck collaboration for the spectral parameters $r, n_s, \alpha_s,
\beta_s$ may be summed up as:
\begin{enumerate}
\item The spectral values provided by PLANCK2013 for $r,n_s,\alpha_s$,
in models without running of the running,
disfavour all of the studied models with a confidence limit
in the ranges 93\%-95.5\%. The poor fit of the value of the running $\alpha_s$
supplied by the models to the measured one, which
has a larger than expected size, is the
most common reason for this invalidation.
\item The corrected values provided by PLANCK2015 for $r,n_s,\alpha_s$,
in models without running of the running, may be fitted by most models
in our study. The main reason for the change is the greatly diminished
measured value of the running $\alpha_s$.
\item  But if one uses the values provided by PLANCK2015, for models
including running of the running, for the parameters $r,n_s,\alpha_s,\beta_s$
again most models are disfavoured with a confidence limit
92.8\% or better. The most common reason is the poor fit these models
provide for the running of the running $\beta_s$, which in this measurement has a larger
than expected size.
\end{enumerate}

The dramatic contrast in results leads to the question of the reliability
of the successive determinations of the spectral parameters $r,n_s,\alpha_s,\beta_s$.
The authors have found that the methodology of model fitting used
in \cite{Ade},\cite{Planck:2015xua} has a bias that often leads to
the overestimation of the highest order parameter, which in turn
triggers biases in the estimation of the lower order parameters.

The bias does not arise from the probabilistic algorithms, such as the
Monte Carlo Markov Chain (MCMC) method,
employed to determine the measured values of the spectral parameters.
It comes from the procedure to judge the fit of the model.

It is another manifestation of the well known bias of the least square
regression polynomial fit when the fitted polynomial $p(x)$ has too small a degree:
the highest order coefficient of $p(x)$ is found to have a very large
value, which is actually caused by the measured data growing faster
than the degree of $p(x)$ allows. The second highest order coefficient
of $p(x)$ typically has a bias of opposite sign to compensate for
the overestimation of the highest degree term. Figure 1
illustrates
this phenomenon with an elementary example.

\medskip

This problem is a particular case of the more general difficulty in asessing the
value of a Taylor polynomial, that approximates a function in the neighbourhood
of a Taylor expansion point, from values of the function at points that lie further and
further from the expansion point. Conceding any weight at all to the values
of the function at points far from the expansion point in the assessment of best
fit carries a great risk of introducing a bias such as that described in Figure 1.

\medskip

The fitting of the model in every step of the MCMC algorithm of \cite{Ade},\cite{Planck:2015xua}, and in comparable algorithms, is probabilistic, but ultimately
close to the classical regression fit: the spectral parameters
on which the model depends are given values that maximize the
likelyhood of the observations that have to be described by the model.
The coefficients $C_l$ of the observed power spectrum are assumed to follow
a Gaussian distribution. This means that the spectral parameters determining
the theoretical power spectrum $\mathcal P_{\mathcal R}(k)$
for scalar perturbations are selected in order to minimize the distance
between the values $C_l$ given by the model and the observed values $\widehat{C_l}$.

This distance is expressed in standard deviations for the probabilistic fitting
that maximizes likelyhood, and in natural units by the classical polynomial
regression fit that
minimizes residue. But the variation of the standard deviation for each
spectral parameter changes little as the parameter varies its value,
so minimizing the distance in natural units or in standard deviations
ends up assigning very similar values to the spectral parameters. Indeed,
this is the reason why probabilistic methods such as MCMC are often
preferred to regression analyses that are more comprehensive but far more
computationally costly and end up reaching a similar result.

A consequence of this equivalence of distances is that likelyhood-maximizing
Bayesian, MCMC \dots methods,
while computationally vastly more efficient than a regression fit, inherit the latter
method's bias, displayed in Figure 1: fitting a model for a function
$\text{ln}\,\mathcal P_{\mathcal R}(k)$ with parameters up to an insufficient degree
will result in an overestimation of the highest order parameter, which
in turn triggers a cascade of estimation errors in the lower order parameters.

The order of the growth of the modelled function
$\text{ln}\,\mathcal P_{\mathcal R}(k)$ (the logarithm of the power spectrum) is currently unknown,
but the pattern of systematically overestimating the highest order spectral
coefficient in its Taylor series, and drastically revising it down as higher coefficients
are incorporated to the model suggest that this bias is indeed happening:
\begin{itemize}
\item The PLANCK2013 determinations of the value of the running $\alpha_s$ in
a $\Lambda$-CDM model without running of the running (table 5 of \cite{Ade})
attribute to $\alpha_s$ expected values ranging from -0.0149 to -0.0094,
\item but the determinations by PLANCK2013 with the same methodology, adding
a running of the running to the model put the expected value of $\alpha_s$
in the range $[0,0.006]$ (one order the magnitude smaller in absolute value
than with only running; sign uncertain due to proximity to zero).
\item The value of the running of the running $\beta_s$ is estimated by PLANCK2013 (table 5 of \cite{Ade})
to lie in the range $[0.017,0.020]$, and by (19) in PLANCK2015 to lie in the
range $[0.025,0.029]$. These values are one order of magnitude greater than
those attributed to the running in these estimations.
\end{itemize}

The values of the spectral parameters $n_s-1,\alpha_s,\beta_s$ will
be reliably known only when higher order terms in the Taylor series
of $\text{ln}\,\mathcal P_{\mathcal R}(k)$ are known, or a method
without this bias is used to fit the models to the data.

\section{Conclusions}

Evidence mounts, both analytical and numerically, that single field slow rolling inflaton
models do not fit well the observations of the CMB radiation. Sophistications
such as multiple fields, or a breakdown in the slow roll regime, or a
completely different paradigm such as the Matter Bounce Scenario ought to be contemplated.

But the value of the spectral parameters $n_s-1,\alpha_s,\beta_s$ of the CMB
radiation is not
yet well established, so we are witnessing the twilight, rather than the death,
of this family of models.

\medskip

In Section \ref{s:analytical} of this work we have found analytically bounds for the
values of the spectral parameters that most of the single field, slow roll inflation models  support. In Section \ref{s:numerical} we have described our MATLAB software package
that finds, for each model given by its potential function $V(\varphi)$,
all possible values of the spectral parameters $n_s-1,\alpha_s,\beta_s$
that the model supports, for a comprehensive list of values of the
field $\varphi$ and of further parameters on which the potential $V(\varphi)$
may depend. There is a deliberate redundancy in the two approaches, and
both have reached the same conclusions for every experimental determination
of the values of the spectral parameters. The numerical testing
software described in this work can be applied to any single field slow roll model
and determination of the spectral parameters beyond those studied here.

\medskip

The conclusions reached by our analytic/numerical assessment of the likelyhood
of single field, slow roll inflation models vary strongly with the different
determinations of the values of the spectral parameters, but there is a clear
pattern in this variation:
\begin{enumerate}
\item the most accurate determinations PLANCK2013 without
running of the running (such as PLANCK2013+WP+BAO) assign to the running
$\alpha_s$ a big value, that rules out most of the models because in them
$\alpha_s$ is a second order expression in the slow roll parameters and
can only be much smaller in magnitude,
\item the determinations PLANCK2015 give a much smaller value to the running
$\alpha_s$, which can be fitted to most of the tested inflaton models,
\item but if the value of the running of the running $\beta_s$ of PLANCK2015
is included in the likelyhood test, it turns out that its determination
by  PLANCK2015 (both TT+lowP and TT,TE,EE+lowP) has such a big magnitude
that again most of the tested inflaton models cannot furnish values
within 1.8 standard deviations of the expected value.
\end{enumerate}

The reason for this pattern of contradictory conclusions seems to
be a mathematical bias in the method that has been used for the computation of the
values of the spectral parameters, which results in a systematic overestimation
of the highest order one. The bias is a migration of, and very close to,
a classical bias of regression (i.e., minimization of residue) fitting:
if one tries to fit a polynomial
of too low degree to values of a function that actually grows faster,
no matter how flawless the procedure for finding the better fit is,
it will result in a polynomial with an exaggerate value for the magnitude of the
leading term, which will result in a cascade of further errors for the
lower coefficients as the table in Fig. 1 illustrates.

The probabilistic (Bayesian, MCMC, \dots) methods currently used to
fit the value of spectral parameters following a Gaussian distribution
in a model actually minimize the residue, expressed in standard deviations
rather than in natural units, and reproduce this bias.

\medskip

The pattern of disproving/validating/disproving of the single field models by successive determinations
of the spectral parameters is not completely symmetrical, because the instances
in which the models are validated are often extreme, narrow choices for the
values of the parameters on which the model depends. Hence the authors'
suspicion that single field slow roll inflation models will ultimately
have to be discarded.

Nevertheless, to rule out inflation models based on their fit to the measured spectral
parameters $r,n_s,\alpha_s,\beta_s$ will not be possible until the value
of these spectral parameters is reliably known, for which more terms in the
Taylor series of $\text{ln}\,\mathcal P_{\mathcal R}(k)$ or, even better, a determination
 procedure free of its current bias, will be required.

\medskip

This investigation has been supported in part by MINECO (Spain), projects MTM2011-27739-C04-01, and MTM2012-38122-C03-01.

\begin{figure}[h]\label{fig:regressio}
\begin{center}
\includegraphics[scale=0.35]{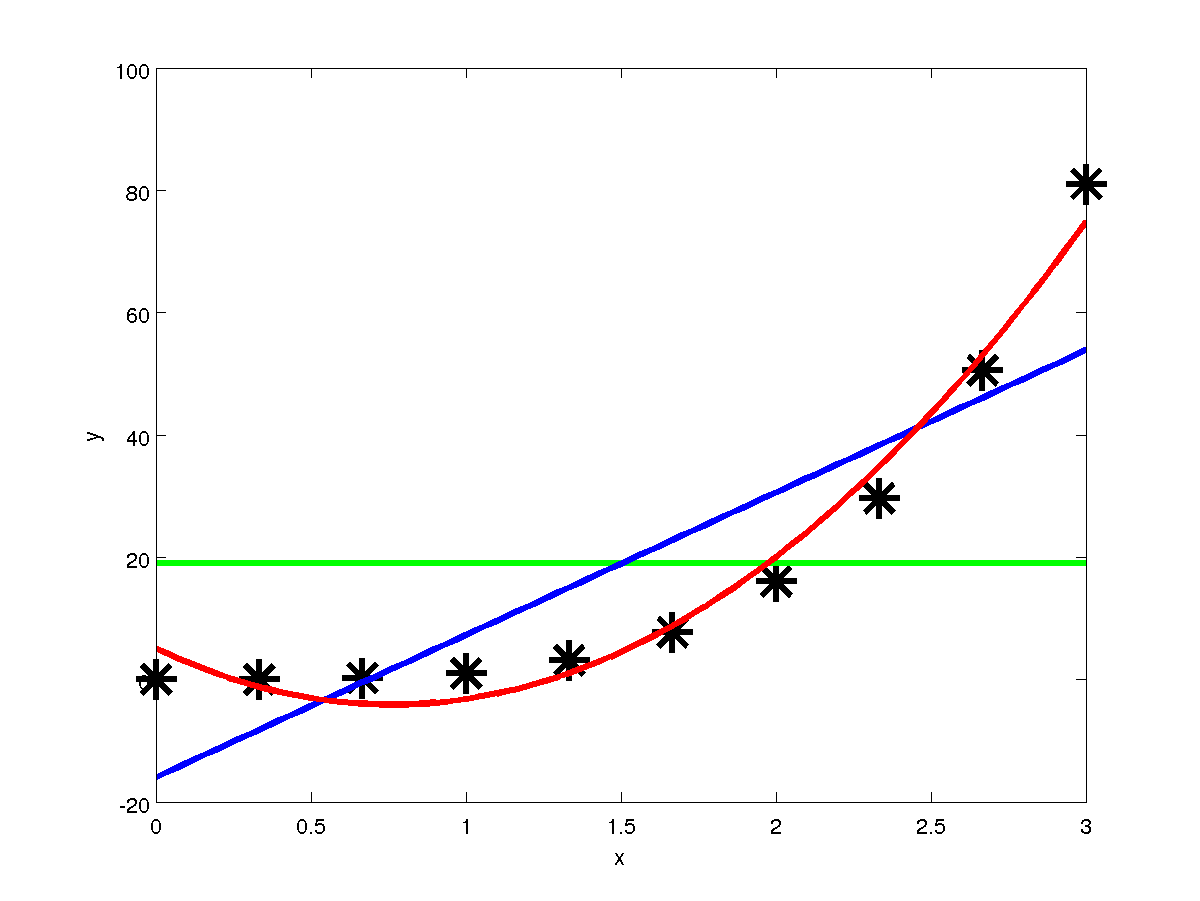}
\end{center}
\caption{
Data points on curve $y=x^4$: the regression polynomials $\sum a_i x^i$ are }
\begin{tabular}{c|ccccc}
degree & $a_0$ & $a_1$ & $a_2$ & $a_3$ & $a_4$ \\
\hline
4 (correct values) & 0 & 0 & 0 & 0 & 1 \\
\hline
3  & -0.53 & 6.67 & -11.22 & 6 &  \\
\hline
2  & 5.07 & -24.07 & 15.78 & & \\
\hline
1  & -15.97 & 23.27 & & & \\
\hline
0  & 18.93 & & & &
\end{tabular}

The leading order coefficient in each regression polynomial is systematically overestimated (from its true
value 0) to try fitting the growth of the function, which is actually of a higher
order. This bias cascades down to the lower order coefficients, starting typically
with an underestimation of the second to highest order coefficient.
\end{figure}

\end{document}